\documentclass{article}

\usepackage{latexsym}
\usepackage{amsmath}
\usepackage{amssymb}
\usepackage{amsthm}
\usepackage{amscd}

\usepackage{graphicx}
\usepackage{subfigure}
\usepackage{psfrag}

\usepackage{bm}
\usepackage{cancel}

\newcommand{\btau}{\bar{\tau}}

\newcommand{\textfrac}[2]{{\textstyle \frac{#1}{#2}}}

\begin{document}


\title{Asymptotics of LRS Bianchi type~I\\cosmological models with elastic matter}


\author{Simone Calogero\footnote{E-Mail: calogero@ugr.es}\\[0.2cm]
Departamento de Matem\'atica Aplicada\\ 
Facultad de ciencias, Universidad de Granada\\
18071 Granada, Spain\\[0.5cm]
J.~Mark Heinzle\thanks{E-Mail: Mark.Heinzle@univie.ac.at}\\[0.2cm]
Gravitational Physics\\ Faculty of Physics, University of Vienna\\
A-1090 Vienna, Austria}






\maketitle


\begin{abstract}


In this paper we report on results in the study of
spatially homogeneous cosmological models with elastic matter.
We show that the behavior of elastic solutions 
is fundamentally different from that of perfect fluid solutions already 
in the case of locally rotationally symmetric (LRS) Bianchi type I models; this is true even when the elastic material
resembles a perfect fluid very closely. In particular, the approach to the initial singularity is characterized by 
an intricate oscillatory behavior of the scale factors, while the future asymptotic behavior
is described by isotropization rates that differ significantly from those of
perfect fluids.


\end{abstract}


\section{Introduction}


In cosmology, the matter model that is used most frequently is that
of a perfect fluid, usually with a linear equation of state. 
This choice is quite general in Friedmann-Robertson-Walker space-times,
where by inheritance of symmetry the stress-energy tensor must have the 
algebraic form of the stress-energy of a perfect fluid.
The case is different, however, for anisotropic space-times: 
Considering perfect fluids is a restriction and
might be misleading, since 
it is unclear 
in general, how robust the results on the behavior of perfect fluid
solutions are under a change of the matter model.
For example, it was shown in~\cite{HU} that the past asymptotic
behavior of Bianchi type I cosmological models with collisionless matter 
is considerably different from that of perfect fluid models. A more systematic analysis of the 
problem in the case of Bianchi type~I models with general anisotropic matter sources has been carried out in~\cite{CH2}.
Most importantly, this difference concerns the behavior 
of solutions in relation to the Taub (flat LRS Kasner) solution.
While there exist perfect fluid solutions that are asymptotic to the Taub solution
as the initial singularity is approached, this behavior does not appear for
solutions with anisotropic matter.
Results of this type have broad ramifications, since the 
dynamics of spatially homogeneous cosmologies are generally conjectured to be the
building blocks of the asymptotic dynamics of 
generic cosmological models (i.e., models without
symmetries), see~\cite{HUR} for a recent discussion.
The Taub solution plays a crucial role in this context;
essential ideas like the basic concept of Kasner eras (cf.\ the BKL asymptotics~\cite{BKL70}) 
and the advanced concept of decay rates of inhomogeneities for generic models are
connected with the Taub solution.
In fact, the Taub solution already plays a crucial role 
in the treatment of Mixmaster dynamics, see~\cite{R}.
It is therefore essential to see in which respect the dynamics
of solutions toward the singularity is dependent on the choice of matter
model, and whether the role of the Taub solution
changes in this context. A systematic analysis 
of this problem in the case of Bianchi type~I cosmological models has been carried out in~\cite{CH2}.

In this paper we widen the analysis of~\cite{CH} 
of the dynamics of cosmological models
with elastic matter; 
the elastic matter model we consider is taken from~\cite{CQ,KS}.
This matter model is particularly suitable for our aims, since it 
contains perfect fluids as a limiting case and thus
allows to directly compare the behavior of anisotropic matter solutions 
(elastic solutions) with the 
behavior of perfect fluid solutions.
In~\cite{CH}, 
it has been shown that 
the asymptotics toward the singularity of elastic solutions
is fundamentally different from that of perfect fluids: The approach to
the singularity is oscillatory, hence, in particular, 
there do not exist any solutions
that approach the Taub solution in the asymptotic limit.
In this paper we study the past and future asymptotic behavior
of elastic models for the locally rotationally symmetric (LRS) Bianchi type~I case in full detail.
In particular, we present a detailed analysis of the oscillatory
behavior of 
the scale factors that determine the metric, 
and we describe the dependence of 
the amplitude of the oscillations
on the properties of the elastic material; see~Section~\ref{pastsection}.
(As we show in~\cite{CH2}, oscillations in the past asymptotics of Bianchi type~I solutions 
are due to an `overcritical' violation of the energy conditions in a neighbourhood of the 
initial singularity; we refer to the concluding remarks.)

In addition to the dynamics toward the initial singularity
we consider the dynamics of elastic cosmologies in the
low density regime, i.e., the future asymptotic behavior of models.
Also in this context we observe a fundamental difference between
elastic solutions and perfect fluid solutions: Although isotropization
occurs for all solutions, the isotropization rates of elastic models 
differ considerably from those of perfect fluid solutions.
This remains true even in the case when the anisotropy properties
of the elastic material are small and the elastic matter thus resembles 
a perfect fluid very closely; 
see~Section~\ref{futuresection}.

The methods we use in this paper are 
methods from the theory of dynamical systems, see, e.g.,~\cite{Perko}.
In Section~\ref{setup} we begin by 
briefly discussing 
the dynamical systems formulation
for Bianchi type~I elastic space-times. 
This approach is
described
in~\cite{CH} in more detail; 
here, however, we adapt the formulation to our present purposes.
Throughout the paper we use unit such that $8 \pi G = 1$ and $c =1$.










\section{Set-up}\label{setup}


Locally rotationally symmetric (LRS) Bianchi type I space-times 
are represented by a line element of the form
\begin{equation}\label{metric}
-d t^2+ A^2 d x^2 + B^2 \left( d y^2+d z^2 \right)\:;
\end{equation}
we denote the spatial part of the metric by $g$.


In the vacuum case,
the solutions of the Einstein equations are the Kasner solutions
\begin{equation}\label{LRSKasnermetric}
-d t^2 + t^{2 a} d x^2 + t^{2 b} \left( d y^2 + d z^2 \right)\:,
\end{equation}
where $a + 2 b = 1 = a^2 + 2 b^2$. There
exist two different solutions,
the non-flat LRS Kasner solution $(a, b) = (-1/3, 2/3)$,
and the Taub solution $(a, b) = (1,0)$.
Since the Taub solution is a representation of a subset of Minkowski space-time,
the Taub metric admits a (smooth) extension 
beyond $t=0$ to Minkowski space-time.
In particular, the hypersurface $t=0$ is null and not
spacelike.

For perfect fluids with
a linear equation of state $p/\rho = w$
three classes of solutions exist.
The Friedmann-Robertson-Walker (FRW) solution is isotropic,



\begin{equation}\label{FRWfluid}
-d t^2 + t^{4/(3[1+w])} \left( d x^2 + d y^2 + d z^2 \right)\,.
\end{equation}
The non-isotropic solutions isotropize 
and approach~\eqref{FRWfluid} as $t\rightarrow \infty$ 
(this will be discussed in detail below);
toward the singularity these solutions are asymptotically
vacuum, i.e., they approach~\eqref{LRSKasnermetric} as $t\rightarrow 0$.
Solutions with $\dot{A}/A < \dot{B}/B$ are asymptotic to
the non-flat LRS Kasner solution;
solutions with $\dot{A}/A > \dot{B}/B$ approach
the Taub solution,
\begin{equation}\label{perfectTaub}
A \propto t \:\Big[ 1 - (1+w)\, t^{1-w}\, \Big] \,,\quad
B \propto 1 + (2-w) t^{1-w}
\end{equation}
as $t\rightarrow 0$. (The proof of~\eqref{perfectTaub} is straightforward,
when one uses the formalism introduced below.)
These solutions possess a
weak null singularity, i.e., like the Taub solution they admit
a (continuous) extension of the metric beyond $t=0$~\cite{R}.
(However, $\rho$ diverges as $t\rightarrow 0$.)


In this paper we consider an LRS Bianchi type I space-time
whose matter content is described by an anisotropic 
stress-energy tensor of the same algebraic type, i.e.,
\begin{equation}\label{matter}
T^\mu_{\ \nu}={\rm diag}\left(\rho,\,p_A,\,p_B,\,p_B\right)\:.
\end{equation}
It is common to define 
the normalized principal pressures,
\begin{equation*}
w_A=\frac{p_A}{\rho},\quad w_B=\frac{p_B}{\rho}\:,
\end{equation*}
the isotropic pressure $p$, and
\begin{equation*}
w=\frac{p}{\rho} = \frac{1}{3} \frac{p_A + 2 p_B}{\rho} = \frac{1}{3} \left(w_A+2w_B\right)\:.
\end{equation*}
For perfect fluids, $w_A = w_B = w$; if this is not the case
the stress-energy tensor is often said to describe an
anisotropic fluid.
However, the required 
specification of the principal pressures 
typically lacks a physical foundation.
A choice of principal 
pressures is usually made \textit{ad hoc} 
to simplify the Einstein equations.









\subsection{Elastic matter}


In this paper we shall consider a stress-energy tensor of the form~\eqref{matter} 
that represents \textit{elastic matter}. 
The general relativistic theory of elasticity has been originally formulated 
by Carter and Quintana in~\cite{CQ} and further elaborated by
by Kijowski/Magli~\cite{KM}, Beig/Schmidt~\cite{BS} 
and Karlovini/Samuelsson~\cite{KS}. Relativistic elasticity is 
used in both relativistic astrophysics and cosmology, see, e.g.,~\cite{KS} and~\cite{Mag}.

An elastic material is specified 
by a constitutive equation (Lagrangian) that depends on scalar functions 
constructed from the configuration map between the space-time and
the natural (unstrained) state of the material;
the stress-energy tensor $T^{\mu}_{\ \nu}$ 
is then obtained as the variation w.r.t.\ the space-time metric of the matter 
action. In particular, this yields expressions for the principal pressures without
the need to resort to any \textit{ad hoc} assumptions;
we refer to~\cite{KS} and 
to the appendix at the end of this paper.
Note in this context that the anisotropies of the elastic stress-energy tensor $T^{\mu}_{\ \nu}$ 
are not due to intrinsic anisotropies
in the elastic matter model, but to anisotropies of the space-time
(provided the unstrained state of the elastic material is assumed
to be isotropic).
More specifically, for the constitutive equation used in~\cite{CQ,KM} the energy density 
is
\begin{equation}\label{rho}
\rho = \rho_0(A B^2)^{-(1+w)} \left( 1 + \frac{v w}{6} \frac{(A^2-B^2)^2}{A^2 B^2} \right)\:,
\end{equation}
where $\rho_0>0$ is a constant, and the (normalized) principal pressures are given by
\begin{subequations}\label{wAB}
\begin{align}
w_A & = w - \frac{v w}{3} \frac{A^4-B^4}{A^2 B^2 + \frac{v w}{6} (A^2-B^2)^2}\,,\\
w_B & = w + \frac{v w}{6} \frac{A^4-B^4}{A^2 B^2 + \frac{v w}{6} (A^2-B^2)^2}\:.
\end{align}
\end{subequations}
These relations contain two constants, $v$ and $w$, where
\begin{equation*}
|w| < 1 \,,\qquad v w > 0 \:
\end{equation*}
is assumed. 
In the appendix of this paper we give a complete derivation of the energy density~\eqref{rho} 
and the principal pressures~\eqref{wAB} for our particular choice of elastic matter and initial data. 

The constant $v$ we call \textit{elastic constant}. It measures the response
of the elastic material to anisotropies; when $v=0$, the elastic
material reduces to a perfect fluid with equation of state $p = w \rho$. 
Therefore, 
the elastic material under consideration contains, as a special case, 
the perfect fluid model most widely used in cosmology, see~\cite{WE}.

By~\eqref{rho}-\eqref{wAB}, there are two regimes in which elastic matter can be 
viewed as a small perturbation of a perfect fluid: When $vw\ll1$ (small shear) or 
when $A\sim B$ (almost isotropic geometry). This observation implies that there are 
indeed regimes where the energy distribution of a perfect fluid and of elastic matter 
are effectively indiscernible. Yet, as we will see, these models give rise to a rather 
different dynamical behavior of the space-time metric in the limits toward the initial singularity and for late times.








\subsection{Dynamical systems formulation}


The Einstein equations for a metric of the type~\eqref{metric} with stress-energy tensor~\eqref{matter}
form a system of ODEs for the variables $(A,\dot{A},B,\dot{B})$; in addition there is one constraint
equation. 
It is preferable, however, to replace $(A,\dot{A},B,\dot{B})$ by scale-invariant variables to obtain 
a dynamical systems formulation of the equations on a compact state-space.
To this end we note that
$\sqrt{g} =  A B^2$\,,
and we define the Hubble scalar as
\begin{equation}\label{Hubblescal}
H = \frac{1}{3} \frac{d}{d t} \left(\log \sqrt{g}\right) = \frac{1}{3} \left[ \frac{\dot{A}}{A} + 2\: \frac{\dot{B}}{B}\right]\:.
\end{equation}
We introduce the dimensionless variables
\begin{equation}\label{pq}
a = \frac{1}{3} \frac{\dot{A}}{A} \, H^{-1}\:,\quad
b = \frac{1}{3} \frac{\dot{B}}{B} \, H^{-1}\:,
\end{equation}
which satisfy the identity $a + 2 b =1$.
In the vacuum case, $(a,b)$ coincide with the constant Kasner parameters and
the metric is~\eqref{LRSKasnermetric};
note, however, that in the general case, $(a,b)$ neither satisfy the relation
$a^2 + 2 b^2 = 1$, nor does the metric assume the form~\eqref{LRSKasnermetric}.
In the variables $(a,b)$ the Hamiltonian constraint equation reads
\begin{equation}\label{Hamcon}
\Omega = \frac{\rho}{3 H^2} = \frac{3}{2} (1 - a^2 - 2 b^2)\:;
\end{equation}
using that $\Omega > 0$ and $a + 2 b =1$ we infer
\begin{equation*}
a \in \big( {-\textfrac{1}{3}}, 1 \big)\:, \quad
b \in \big( 0, \textfrac{2}{3} \big)
\end{equation*}
as well as $\Omega \leq 1$; in addition, $\Omega = 1$ iff
$a = 1/3 =b$, which is the case iff $\dot{A}/A = \dot{B}/B$, i.e.,
for the FRW solution.

As the second set of dimensionless variables we use
\begin{equation}\label{alphabeta}
\alpha = \frac{A^2}{A^2 + B^2} \:,\quad 
\beta = \frac{B^2}{A^2 + B^2} \:,
\end{equation}
where $\alpha + \beta = 1$.
Finally we introduce
a new time variable $\tau$ by defining
\begin{equation}\label{tandtau}
\partial_\tau=H^{-1}\partial_t
\end{equation}
and we denote by a prime the differentiation w.r.t.\ $\tau$.

Using these variables the Einstein equations 
decouple into the dimensional equation
for the Hubble variable 
\begin{equation}
\label{equationH}
H' = -3 H \left[1-\frac{\Omega}{2}(1-w)\right]\:,
\end{equation}
and a reduced system of dimensionless equations
\begin{subequations}
\label{dynamicalLRS}
\begin{align}
\alpha^\prime & = 9 \alpha (1- \alpha) \Big(a -\frac{1}{3} \Big)\:, \\
a^\prime & = -\Omega \left[ \frac{3}{2} ( 1- w) \Big( a-\frac{1}{3} \Big) - (w_A -w )\right]\:,
\end{align}
\end{subequations} 
where $\Omega$ is determined by the constraint~\eqref{Hamcon},
\begin{equation*}
\Omega = \frac{9}{4} (1 - a) \Big(a + \frac{1}{3} \Big)
\end{equation*}
and $w_A$ is given by~\eqref{wAB} expressed in the new variables,
\begin{equation}
\label{w}
w_A  = w + \frac{v w}{3} \frac{1 - 2\alpha}{\alpha (1-\alpha) + \frac{v w}{6} (1-2 \alpha)^2} \:.
\end{equation}
The phase space associated with the dynamical system~\eqref{dynamicalLRS} 
is
\begin{equation*}
\mathcal{L} = \Big( {-\frac{1}{3}}, 1\Big) \times (0,1) \,\ni\, (a, \alpha)\:;
\end{equation*}
since~\eqref{dynamicalLRS} admits a smooth extension to
$\overline{\mathcal{L}}$, it is beneficial to 
include the boundary $\partial\mathcal{L}$ in our analysis.










\section{Dynamical systems analysis}


The boundary $\partial \mathcal{L}$ of the state space 
can be represented as a rectangle.
The four vertices are fixed points of the dynamical system~\eqref{dynamicalLRS}; we denote
these points by
\begin{subequations}
\begin{align}
& \mathrm{Q}_1: (a,\alpha) = (-\textfrac{1}{3},1) & & \mathrm{T}_1: (a,\alpha) = (1,1) \\
& \mathrm{Q}_0: (a,\alpha) = (-\textfrac{1}{3},0) & & \mathrm{T}_0: (a,\alpha) = (1,0)\:.
\end{align}
\end{subequations}
The four sides of the rectangle (where we exclude the fixed points) 
are invariant subsets.
When $a = -1/3$ we find $\alpha^\prime < 0$;
$a =1$ entails $\alpha^\prime > 0$; 
for $\alpha = 1$ we find by using~\eqref{w} that
\begin{equation*}
a^\prime\big|_{\alpha =1} = - \frac{3}{2} \Omega (1-w) \Big[ a + \frac{1+w/3}{1-w} \Big] < 0  \:;
\end{equation*}
analogously, $a^\prime > 0$ when $\alpha =0$.
It follows that the flow on $\partial\mathcal{L}$ forms a heteroclinic cycle:
\begin{equation}\label{heterocycle}
\begin{CD}
\mathrm{Q}_1 @<<< \mathrm{T}_1 \\
@VVV @AAA \\
\mathrm{Q}_0 @>>> \mathrm{T}_0 
\end{CD}
\end{equation}

The solutions associated with the fixed points $\mathrm{T}_0$, $\mathrm{T}_1$, and the 
orbit $\mathrm{T}_0 \rightarrow \mathrm{T}_1$ can be interpreted
as the Taub solution. 
This is because $a = 1$ ($b= 0$) yields
$H \propto 1/(3 t)$ via~\eqref{tandtau} and~\eqref{equationH},
and accordingly,~\eqref{pq} leads to $A \propto t$ and $B = \mathrm{const}$, i.e.,
to the Taub solution.
Analogously, the fixed points $\mathrm{Q}_0$, $\mathrm{Q}_1$, and the
orbit $\mathrm{Q}_0 \leftarrow \mathrm{Q}_1$ are representations
of the non-flat LRS Kasner solution, since $a = -1/3$ (and thus $b = 2/3$).

In the interior of 
$\mathcal{L}$ 
there is one fixed point:
\begin{equation*}
\mathrm{F}: (a,\alpha) = (\textfrac{1}{3},\textfrac{1}{2}) \:.
\end{equation*}
The solution associated with $\mathrm{F}$
is isotropic, since $a = 1/3$ (and thus $b = 1/3$, $\Omega= 1$);
$\alpha=\beta =1/2$ entails $w_A = w_B = w$.
Accordingly, the fixed point $\mathrm{F}$ represents
the FRW perfect fluid solution~\eqref{FRWfluid}, which
is associated with a perfect fluid with equation of 
state $p = w \rho$.

\begin{figure}[Ht]
\begin{center}
\psfrag{yj}[cc][cc][1][0]{$\alpha$}
\psfrag{sigj}[cc][cc][1][0]{$a$}
\psfrag{Qkij}[cc][cc][1][0]{$\text{T}_1$}
\psfrag{Qjik}[cc][cc][1][0]{$\text{T}_0$}
\psfrag{Tjik}[cc][cc][1][0]{$\text{Q}_0$}
\psfrag{Tkij}[cc][cc][1][0]{$\text{Q}_1$}
\psfrag{13}[cc][cc][1][0]{$1/3$}
\psfrag{F}[cc][cc][1][0]{{\bf F}}
\includegraphics[width=0.7\textwidth]{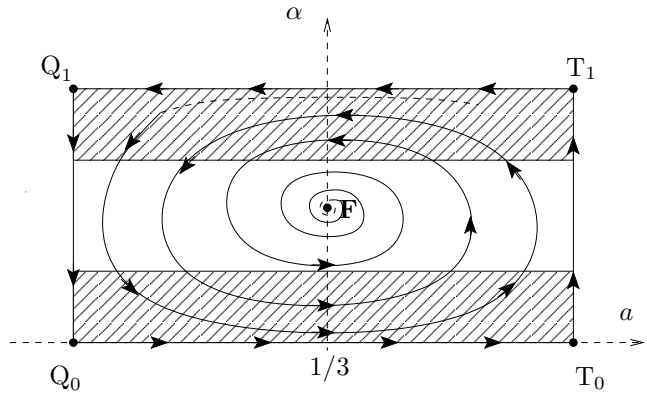}
\caption{A generic orbit in the phase space $\mathcal{L}$. 
It is assumed that $v w >(3/32) (1 -w)^2$. Under this assumption
we observe oscillatory approach toward the FRW solution represented by $\mathrm{F}$.
The dominant energy condition is violated in the shadowed region.}
\label{lrspicture}
\end{center}
\end{figure}

To analyze the global dynamics of the system~\eqref{dynamicalLRS} on $\mathcal{L}$
we introduce the function
\begin{equation*}
Z = (1-a)^{-1} \Big( a + \frac{1}{3} \Big)^{-1} \left[ 1 + \frac{v w}{6} \frac{(1- 2 \alpha)^2}{\alpha(1-\alpha)} \right]\:,
\end{equation*}
which is positive on $\mathcal{L}$; in fact, 
$Z = 1 = \inf_\mathcal{L} Z$ at $\mathrm{F}$,
$Z > 1$ on $\mathcal{L} \backslash \{\mathrm{F}\}$,
and $\sup_{\mathcal{L}} Z=+\infty= Z|_{\partial\mathcal{L}}$. 
A straightforward calculation shows that $Z$ is decreasing along 
all orbits (different from $\mathrm{F}$), 
i.e., $Z^\prime < 0$
on $\mathcal{L}\backslash \{\mathrm{F}\}$.
The monotonicity principle~\cite{FHU, WE}
implies that (i) each orbit in $\mathcal{L}$ 
converges to $\mathrm{F}$ as $\tau\rightarrow \infty$;
(ii) each orbit (different from $\mathrm{F}$) 
approaches $\partial\mathcal{L}$ as $\tau\rightarrow -\infty$.
Since $\partial\mathcal{L}$ is given by~\eqref{heterocycle},
the past asymptotic behavior of solutions is represented
by this heteroclinic cycle.
This behavior of solutions is depicted in Figure~\ref{lrspicture}.

Equations~\eqref{tandtau} and~\eqref{equationH} allow us to translate 
between $\tau$-time and synchronous time $t$.
Since $-3 H \leq H^\prime \leq -3 (1 +w) H/2$, Equation~\eqref{tandtau} can be
integrated to yield a positive function $t(\tau)$ that
satisfies 
$t\searrow 0$
as $\tau\rightarrow -\infty$.

Thus the interpretation of the results on the global dynamics of solutions
is the following:
Each LRS Bianchi type I model with elastic matter 
isotropizes toward the future (i.e., as $t\rightarrow \infty$)
and (to first order) 
behaves like an (infinitely diluted) isotropic perfect fluid
solution in the asymptotic regime. 
(At higher orders, however, when we consider isotropization rates,
the behavior of elastic solutions differs significantly from that
of perfect fluid solutions; see below.)

Toward the singularity, i.e., as $t\rightarrow 0$,
we observe \textit{oscillatory} behavior of elastic cosmologies.
Elastic models do not converge to either the Taub solution
or the non-flat LRS solution, but exhibit oscillations between
the two, which is a direct consequence of 
the approach of solutions to the heteroclinic cycle~\eqref{heterocycle}.

In the following we investigate the future and past asymptotics
of elastic models in detail.


\section{Future asymptotics and isotropization rates}\label{futuresection}





The analysis of the future asymptotic dynamics 
is based on an investigation of the dynamical system~\eqref{dynamicalLRS} in the neighborhood
of the equilibrium point $\mathrm{F}$.
The linearization of~\eqref{dynamicalLRS} at $\mathrm{F}$
possesses the eigenvalues $\lambda_{1}$, $\lambda_{2}$
and associated eigenvectors $v_1$, $v_2$,
\begin{align*}
\lambda_{1,2} & = \frac{3}{4} \left( -(1-w) \mp \sqrt{(1-w)^2 - \frac{32}{3}\: v w}\right)\,,\\
v_{1,2} & = \left(\frac{v w}{3}\,,\, \frac{3}{16} \,
\left[(1-w) \mp \sqrt{(1-w)^2 -\frac{32}{3}\:v w} \right]\right)^{T}\:.
\end{align*}
It is immediate that, if $v w \leqslant \frac{3}{32}(1-w)^2$, the eigenvalues 
of the linearization of the system at $\mathrm{F}$ 
are real and negative; in this case
$\mathrm{F}$ is a stable node.
If $v w > \frac{3}{32}(1-w)^2$
the eigenvalues are complex (with negative real part); in this case
the fixed point $\mathrm{F}$ is a stable focus (stable spiral)
and the solutions' approach to $\mathrm{F}$ as $\tau\rightarrow \infty$
is oscillatory; see Figure~\ref{lrspicture}.
The late time behavior of elastic cosmologies 
is thus characterized by
\begin{itemize}
\item[(i)] monotonic isotropization if $v w \leqslant (3/32)(1-w)^2$;
\item[(ii)] oscillatory isotropization if $v w > (3/32)(1-w)^2$.
\end{itemize}

To compare the behavior of elastic models with the behavior
of the associated perfect fluid solutions
we consider elastic matter that behaves ``almost like a perfect fluid'',
i.e., we choose $v$ to be small,
\begin{equation}\label{absmall}
v w \ll (1-w)^2\:.
\end{equation}
Evidently, this assumption implies monotonic isotropization of solutions.
It follows from~\eqref{absmall} that
\begin{align*}
\lambda_1 & = -\frac{3}{2} \:(1-w)\:, &
\lambda_2 & = -4 \:\frac{v w}{1-w} \:,\\
v_1 & = \begin{pmatrix} \frac{1}{3}\:(1-w)  \\ 1 \end{pmatrix} \:,&
v_2 & = \begin{pmatrix}  v w \\ \frac{9}{8}\:(1-w)\end{pmatrix}\:,
\end{align*}
in the lowest order approximation.
Accordingly, the solutions of the dynamical system in the
neighborhood of $\mathrm{F}$ are approximately given by














\begin{equation}\label{solapprox}
\begin{pmatrix}
a\\
\alpha 
\end{pmatrix} 
=
\begin{pmatrix}
\frac{1}{3}\\
\frac{1}{2}
\end{pmatrix}
+ c_1 v_1 \,e^{-|\lambda_1| \tau}\\
+ c_2 v_2 \,e^{-|\lambda_2| \tau}\:.
\end{equation}
Consider first the solution determined by $c_1 \neq 0$, $c_2 = 0$.
This solution coincides 
with the general \textit{perfect fluid solution},
i.e., the solution to the (decoupled) equations
\begin{equation*}
a^\prime =  -\frac{3\Omega}{2} ( 1- w ) \Big( a-\frac{1}{3} \Big) \:,\quad
\alpha^\prime = 9 \alpha (1- \alpha) \Big(a -\frac{1}{3} \Big)\:, 
\end{equation*}
which are obtained from~\eqref{dynamicalLRS} by setting $w_A=w_B \equiv w$
and thus form the system of equations for LRS Bianchi type I perfect fluid models
(satisfying the equation of state $p = w \rho$).
Integrating~\eqref{equationH} we obtain
\begin{equation}
H \propto e^{-\frac{3}{2} (1+w) \tau} \left(1 + \mathrm{const}\: e^{-3 (1-w) \tau}\right)\:,
\end{equation}
which is subsequently used in Equation~\eqref{tandtau} to yield
the functional relation between $\tau$-time and 
synchronous time $t$:
\begin{equation}\label{ttau}
t \propto e^{\frac{3}{2} (1+w) \tau} \left( 1 + \mathrm{const} \: e^{-3 (1-w)\tau} \right)\:.
\end{equation}
(In the case $w=1/3$ the lower order term is of a different form.)





Finally, using $(\sqrt{g})^\prime = 3 \sqrt{g}$, which follows from~\eqref{Hubblescal} and~\eqref{tandtau}, 
we obtain
\begin{subequations}\label{nongeneric}
\begin{align}
\sqrt{g} & = \sqrt{g_0} \, t^{\frac{2}{1+w}} \left( 1 + \mathrm{const}\: t^{-2 \frac{1-w}{1+w}} \right)\:, \\
A(t) & = (\sqrt{g_0})^{1/3} \, t^{\frac{2}{3} \frac{1}{1+w}} \left( 1 - 2 \,\mathrm{const} \:  t^{-\frac{1-w}{1+w}}\right)\:, \\
B(t) & = (\sqrt{g_0})^{1/3} \, t^{\frac{2}{3} \frac{1}{1+w}} \left( 1 + \mathrm{const} \:  t^{-\frac{1-w}{1+w}}\right)\:.
\end{align}
\end{subequations}
However, it is straightforward to see from~\eqref{solapprox} that this solution 
is \textit{not generic} among LRS Bianchi type I solutions with elastic matter.
Since $|\lambda_1| \gg |\lambda_2|$, in~\eqref{solapprox}, the term 
$c_1 v_1 e^{-|\lambda_1| \tau}$ 
can be neglected with respect to $c_2 v_2 e^{-|\lambda_2| \tau}$, and thus
the generic solution behaves asymptotically
like a solution given by $c_1 = 0$, $c_2 \neq 0$.















Proceeding in complete analogy to above we find 
\begin{subequations}\label{generic}
\begin{align}
\sqrt{g} & = \sqrt{g_0} \, t^{\frac{2}{1+w}} \left( 1 + \mathrm{const}\: t^{-\frac{16}{3} \frac{v w}{1-w^2}} \right) \:,\\
A(t) & = (\sqrt{g_0})^{1/3} \, t^{\frac{2}{3} \frac{1}{1+w}} \left( 1 - 2 \,\mathrm{const} \:  t^{-\frac{8}{3} \frac{v w}{1-w^2}}\right)\:, \\
B(t) & = (\sqrt{g_0})^{1/3} \, t^{\frac{2}{3} \frac{1}{1+w}} \left( 1 + \mathrm{const} \:  t^{-\frac{8}{3} \frac{v w}{1-w^2}}\right)\:,
\end{align}
\end{subequations}
for the generic elastic model.
Comparing~\eqref{nongeneric} and~\eqref{generic} we observe a fundamental difference
in the \textit{isotropization rates}
which are defined as 








\begin{equation*}
\mathrm{iso}(t)  = (\sqrt{g})^{-1/3} \left|  A - (\sqrt{g})^{1/3} \right| \:.
\end{equation*}


\begin{itemize}
\item[(I)] There exists one single solution to the LRS Bianchi type I equations
with elastic matter that behaves (to second order) like a perfect fluid as $t\rightarrow \infty$. 
In particular, the
isotropization rates are identical:
\begin{equation}\label{isofl}
\mathrm{iso}(t) \, \propto\, t^{-\mathrm{iso}_{\mathrm{f}\mathrm{l}}} = t^{-\frac{1-w}{1+w}} \:.
\end{equation}
\item[(II)] For the generic solution 
the isotropization rate is not the one of a perfect fluid solution, but
\begin{equation}\label{isoel}
\mathrm{iso}(t) 
\,\propto \, t^{-\mathrm{iso}_\mathrm{el}} =t^{-\frac{8}{3} \frac{v w}{1-w^2}} \:.
\end{equation}
\end{itemize}
In particular, since $\mathrm{iso}_\mathrm{el} \ll \mathrm{iso}_{\mathrm{f} \mathrm{l}}$,
isotropization occurs at a much slower rate.

We see that, therefore, 
the generic late-time behavior of (non-isotropic) solutions with elastic matter
is considerably different from the behavior of (non-isotropic) perfect
fluid solutions. Since this is despite the fact that the material properties 
of the elastic matter resemble those of a perfect fluid,
this result is interesting.

In astrophysical applications, such as the 
modeling of relativistic stars, replacing perfect fluid matter by elastic matter 
yields a family of models that includes the perfect fluid solutions
as special cases (which are obtained by simply letting the elastic constants go to zero, see~\cite{KS}).
Here we see that, in contrast, in (spatially homogeneous) cosmology, 
elastic models behave qualitatively different from fluid models.









\section{Past asymptotics}\label{pastsection}


In the asymptotic regime $\tau\rightarrow -\infty$ (i.e., $t\rightarrow 0$), 
every solution of~\eqref{dynamicalLRS} is described by 
the heteroclinic cycle~\eqref{heterocycle} to an increasing degree of accuracy.
Accordingly, we observe alternations between Taub phases
and non-flat LRS phases:
In Taub phases, which are associated
with the orbit in Figure~\ref{lrspicture} being close to 
either of the fixed points $\mathrm{T}_0$, $\mathrm{T}_1$ or
the orbit $\mathrm{T}_0\rightarrow \mathrm{T}_1$ connecting these points,
the solution is approximated 
by a Taub solution, i.e., 
\begin{subequations}
\begin{align}
\label{flatphase}
& A \propto t \:, \qquad  B \propto\mathrm{const}\:;
\intertext{in non-flat LRS phases, which are associated
with the orbit being close to 
either of the fixed points $\mathrm{Q}_0$, $\mathrm{Q}_1$ or
the orbit $\mathrm{Q}_0\leftarrow \mathrm{Q}_1$, the solution is approximated by}
\label{nonflatphase}
& A \propto t^{-1/3}\,, \quad  B \propto t^{2/3}\:.
\end{align}
\end{subequations}
Elastic cosmological models will oscillate between 
phases~\eqref{flatphase} and~\eqref{nonflatphase} 
with a rapidly increasing frequency as $t\rightarrow 0$.
This is a simple consequence of our considerations.

A priori, oscillations between phases of the type~\eqref{flatphase} and~\eqref{nonflatphase}
are compatible with a large variety of scenarios: 
There could exist any constants $0 \leq c_1 \leq c_2 \leq \infty$
such that $\liminf_{t\rightarrow 0} A = c_1$ and $\limsup_{t\rightarrow 0} A = c_2$.
In the following, however, we show that 






\begin{equation}\label{Aclaim}
\begin{split}
\text{(i)}\quad &\lim_{t\rightarrow 0} A = 0 \,, \qquad\qquad\qquad\qquad\quad\: \:\text{if } w>0\:; \\
\text{(ii)}\quad &\liminf_{t\rightarrow 0} A = 0,\:\: \limsup_{t\rightarrow 0} A = \infty \,,
\quad   \text{if } w<0\:.
\end{split}
\end{equation}

Accordingly, the amplitude of the oscillations converges to zero
if $w>0$, and diverges if $w<0$.
Note that convergence of the scale factor $A$ in the case $w>0$ occurs
despite the fact that there exist phases~\eqref{nonflatphase} where $A$ is increasing.

In order to prove the asserted asymptotic properties of the scale factor $A$,
we analyze in detail the asymptotic behavior of orbits in $\mathcal{L}$.
Let $\epsilon > 0$; we define an $\epsilon$-neighborhood
of a point $(\mathring{a}, \mathring{\alpha})\in\overline{\mathcal{L}}$ 
as the set of all points $(a,\alpha)$ 
such that $|\alpha-\mathring{\alpha}| \leq \epsilon$ and $|a-\mathring{a}| \leq \epsilon$.
If $\epsilon$ is sufficiently small, then the flow of
the dynamical system~\eqref{dynamicalLRS} in an $\epsilon$-neighborhood
of a fixed point can be approximated by the linearized system.

Consider an arbitrary orbit $\gamma$ in $\mathcal{L}$. 
To facilitate matters we invert the direction of time by
defining $\btau = -\tau$; hence the orbit
$\gamma(\btau) = (a,\alpha)(\btau)$ approaches the heteroclinic cycle~\eqref{heterocycle}
as $\btau \rightarrow \infty$ (instead of $\tau\rightarrow -\infty$).
There exists an increasing sequence 
of times $\btau_n$, $n\in \mathbb{N}$,
such that $\gamma(\btau_n) = (a,\alpha)(\btau_n) = (1-\epsilon, \alpha_n)$, where $\alpha_n < \epsilon$ 
for all $n$.
Our proximate aim is to analyze the sequences 
\begin{equation}
(\btau_n)_{n\in\mathbb{N}} \quad\text{and}\quad (\alpha_n)_{n\in\mathbb{N}}\:.
\end{equation}
By construction, at time $\btau_n$, the orbit $\gamma$ enters the $\epsilon$-neighborhood of 
the fixed point $\mathrm{T}_0$. 
Using the linearized flow it is straightforward
to compute that $\gamma$ leaves this neighborhood again
at time $\btau_n + \Delta\btau$, where $\Delta\btau = (1/6) \log (\epsilon/\alpha_n)$; 
furthermore,
$\gamma(\btau_n + \Delta\btau)$ is given by $\big(1 - \epsilon^{-(1-w)/2} \alpha_n^{(3-w)/2}, \epsilon\big)$. 
Subsequently, the orbit $\gamma$ is approximated by a linear perturbation
of the boundary orbit $\mathrm{T}_0\rightarrow \mathrm{T}_1$.





The straightforward calculations show that 
the passage from the $\epsilon$-neighbor\-hood of $\mathrm{T}_0$ to
the $\epsilon$-neighborhood of $\mathrm{T}_1$ takes 
the time $\Delta\btau = -(1/3) \log\epsilon$ and that
the orbit $\gamma$ enters
the $\epsilon$-neighborhood of $\mathrm{T}_1$ at a point with coordinates
$\big(1 - c_{\mathrm{T}}\,\epsilon^{-(1-w)/2} \alpha_n^{(3-w)/2}, 1-\epsilon\big)$
where $c_{\mathrm{T}}$ is a positive constant that depends on the choice of
$\epsilon$ only. Proceeding in this way, i.e., by tracking the orbit $\gamma$ through the
$\epsilon$-neighborhoods of the fixed points and along the boundary orbits,
we obtain straightforwardly that
\begin{equation}\label{alphan}
\alpha_{n+1} = C\, \epsilon \, \left( \frac{\alpha_n}{\epsilon} \right)^{(3-w)^2/(1+w)^2}\:,
\end{equation}
where $C$ is a positive constant that depends on $\epsilon$ (but is independent of $n$).
As a alternative to~\eqref{alphan} we can write
\begin{equation}\label{alphanalt}
\left(\frac{\alpha_{n+1}}{\tilde{\epsilon}}\right) = \left( \frac{\alpha_n}{\tilde{\epsilon}} \right)^{(3-w)^2/(1+w)^2}\:,
\end{equation}
where $\tilde{\epsilon}$ is a positive constant (independent of $n$).
Iteration yields
\begin{equation}
\left(\frac{\alpha_{n}}{\tilde{\epsilon}}\right) = \left( \frac{\alpha_0}{\tilde{\epsilon}} \right)^{(3-w)^{2n}/(1+w)^{2n}}\:.
\end{equation}
Note that by choosing $\alpha_0 < \tilde{\epsilon}$ we achieve $\alpha_0/\tilde{\epsilon} < 1$.
Similarly, for $\btau_{n+1} - \btau_n$ we find
\begin{equation}
\btau_{n+1} = \btau_n -\frac{8}{3 (1+w)^2} \log\frac{\alpha_n}{\tilde{\epsilon}} + \mathrm{const}\:,
\end{equation}
and therefore
\begin{equation}
\btau_n - \btau_0 = 
-\frac{1}{3 (1-w)} \log \frac{\alpha_{n}}{\tilde{\epsilon}} \left( 1 + O(n C^{-n})\right)\:,
\end{equation}
where $C > 1$, or, approximately,
\begin{subequations}\label{alphanalphaprimen}
\begin{equation}\label{alphanres}
\alpha_n = \mathrm{const}\: e^{-3 (1-w) \btau_n} \:.
\end{equation}

In a completely analogous manner one can find a different sequence
of times, which we denote by $\btau^\prime_n$, 
and a associated sequence $\gamma(\btau^\prime_n) = (1-\epsilon, \alpha^\prime_n)$,
where $\alpha^\prime_n = \alpha(\btau^\prime_n)$,
such that
\begin{equation}\label{alphaprimenres}
\alpha_n^\prime = 1 -  \mathrm{const}\: e^{-3 (1-w) \btau^\prime_n}\:.
\end{equation}
\end{subequations}

By construction, $\alpha_n$ is related to the minimum of $\alpha$ in 
the interval $[\btau_{n-1}, \btau_n]$ (which corresponds to 
one oscillation). To see that, note that
the minimum is attained somewhere along the orbit $\mathrm{Q}_0 \rightarrow \mathrm{T}_0$;
using again the linearized flow in the neighborhood of that
boundary orbit, we infer that $\min_{\btau \in [\btau_{n-1}, \btau_n]} \alpha (\btau) = k \alpha_n$ for
some constant $k$ independent of $n$.
Analogously, $\alpha_n^\prime$ is related to 
the maximum of $\alpha$, i.e., 
$\max_{\btau \in [\btau^\prime_{n-1}, \btau^\prime_n]} \alpha(\btau) = k^\prime \alpha^\prime_n$
for some constant $k^\prime$.

Since 
\begin{equation}
A^3 = \sqrt{g} \,\frac{\alpha}{1-\alpha} 
\end{equation}
by~\eqref{alphabeta},
and $\sqrt{g} = \sqrt{g_0} e^{-3 \btau}$ by~\eqref{Hubblescal} and~\eqref{tandtau},
it follows that
\begin{subequations}\label{AAprime}
\begin{align}
A_n & = A(\btau_n) \,\propto\, \exp\left[ -(2-w) \btau_n\right]\,,\\[1ex]
A^\prime_n & = A(\btau^\prime_n) \,\propto\, \exp\left[-w \btau^\prime_n\right]\:.
\end{align}
\end{subequations}
Consider the case $w>0$. Since $A^\prime_n$ is a measure
for the maximum that $A$ attains during one oscillation period
(and $A_n$ a measure for the minimum), we find
that $A(\btau) \rightarrow 0$ as $\btau \rightarrow \infty$;
in fact, \eqref{AAprime} can be condensed into the statement that
\begin{equation}\label{Abound}
A(\btau) \:\leq\: \mathrm{const}\: e^{-w \btau}
\end{equation}
as $\btau\rightarrow \infty$ (or $\tau \rightarrow -\infty$, $t\rightarrow 0$).
Note incidentally that $e^{-\btau} \propto t^{1/3}$ as $t\rightarrow 0$; this can be shown
via~\eqref{equationH} and a line of reasoning analogous to the above. 
We have thus proved the claim that, in the case $w >0$, the scale factor $A$ converges to zero;
this convergence occurs despite the existence of phases~\eqref{nonflatphase} where $A$ is increasing. 
In the case $w < 0$, the conclusion from~\eqref{AAprime} is that the amplitudes
of the oscillations grow unboundedly in such a way that 
$\liminf_{t\rightarrow 0} A = 0$ and $\limsup_{t\rightarrow 0} A = \infty$.
This concludes the proof of the claim~\eqref{Aclaim}.

Interestingly enough, the behavior of
the scale factors is largely determined by the constant $w$,
while the dependence on the elastic constant $v$ is minor: It is hidden
in the constants appearing in~\eqref{alphanalphaprimen} and in 
the derived formulas~\eqref{AAprime} and~\eqref{Abound}.
To explain the quasi-independence of 
the qualitative asymptotic behavior of the scale factors
of the elastic constant $v$, 
we note that, in the asymptotic regime,
the orbit spends a rapidly increasing amount of $\tau$-time
in the neighborhood of the fixed points,
while the time that elapses during  
the passage from one fixed point to the other is fixed.
Since the fixed points thus dominate the asymptotic evolution of solutions
and since the flow in the neighborhood of the fixed points
is independent of $v$, it is only the isotropic constant $w$ that enters in the description
the qualitative asymptotic behavior of the scale factors.

Summarizing we see that the past asymptotic behavior of elastic cosmologies,
as described by~\eqref{AAprime} and~\eqref{Abound},
is in stark contrast to the behavior of perfect fluid
solutions, which converge to either the Taub solution (when $a > b$)
or to the non-flat LRS solution (when $a < b$) 
as $t\rightarrow 0$; see~\eqref{perfectTaub}.
The structure of the singularity is therefore completely different
for elastic models. In particular, there does not
exist the option of a weak null singularity.














\section{Discussion and conclusions}


In this paper we analyze locally rotationally symmetric (LRS)
models of Bianchi type~I with elastic matter.
Since the elastic matter model naturally contains perfect fluid matter as
a limiting case---the latter being the matter model
most commonly used in cosmology---,
we are able to directly compare the
behavior of elastic models with the behavior of perfect fluid cosmologies.

Toward the future all elastic models approach an infinitely dilute isotropic
state.
The approach to this state is oscillatory when the elastic constant
is sufficiently large; when the constitutive equation of state of the elastic matter does not deviate considerably from that of a perfect fluid
(i.e., when the modulus of rigidity is small) isotropization
is monotonic.
Interestingly enough, even in the latter case, 
isotropization occurs at a rate that is fundamentally different
from the isotropization rates observed for perfect fluids models.
This is in contrast to standard astrophysical applications where elastic materials 
produce models that closely display the perfect fluid behavior~\cite{KS}. 
In particular perfect fluid solutions are recovered by letting the elastic constant
$v\to 0$. In the cosmological context this is no longer true; 
as shown in this paper, the isotropization rates of elastic solutions are qualitatively  different from those of perfect fluid solutions

The past asymptotics of elastic models is also interesting.
While perfect fluid solutions converge to either the Taub
solution or the non-flat LRS solution, elastic models
oscillate between these two states. Oscillatory behavior toward the initial singularity
is well-known in the context of certain Bianchi cosmologies,
most notably in Bianchi types~VIII and~IX. 
This asymptotic behavior is usually referred to as the Mixmaster
behavior.
However, Mixmaster oscillations 
are absent when LRS symmetry is imposed: LRS solutions 
simply approach the Taub solution or the non-flat LRS solution.
This is in stark contrast to behavior of elastic models
analyzed in the present paper. These models exhibit
oscillations despite the fact that they are LRS models.

Elsewhere we show in more detail that there is no evident connection between
elastic oscillations and Mixmaster oscillations~\cite{CH}.
(It might be justified to say that the former are caused by the matter
and the latter by geometry).
In the paper~\cite{CH} we consider the case of generic (in fact, diagonal) Bianchi type~I models,
and we find an intricate network of oscillations that determine
the past asymptotic behavior of these elastic cosmologies.
Again, this chaotic oscillatory approach
to the singularity is fundamentally different from the
Mixmaster behavior.

Despite the fact that elastic oscillations are not directly connected with
Mixmaster oscillations, there might exist interesting
consequences when one considers more general elastic models
than Bianchi type~I.
Consider, e.g., elastic models of Bianchi type~VIII or~IX, say.
In this case we expect Mixmaster behavior, which 
is characterized by oscillations between epochs where
the behavior is close to the behavior of Bianchi type~I models.
Since already the Bianchi type~I models are oscillatory,
as shown here and in~\cite{CH}, we might be confronted with
a hierarchy of oscillations, i.e.,
oscillations between oscillations, where Mixmaster oscillations 
connect epochs of elastic oscillations.
Whether this is indeed true remains to be investigated.

Although 
it is not the purpose of this paper to propose elastic matter as a 
physically realistic matter model for the universe,
we would like to conclude this paper by commenting on the `physics' of elastic matter
and elastic cosmologies.
We have seen that
the asymptotic behavior toward the initial singularity of elastic cosmological
models differs qualitatively (and not merely quantitatively)
from that of perfect fluid models.
This is hardly surprising, since
the differences in the material properties of 
elastic matter and perfect fluid matter 
are most significant 
under extreme conditions, such as those found in the proximity of 
a singularity.
One could expect that 
a description of matter as a solid elastic body might not be completely unrealistic
under these conditions---elastic matter is also used
in the modeling of compact stellar objects like neutron stars.
However, we note that 
the dynamics toward the singularity of the elastic cosmologies under study are
connected with the violation of energy conditions.
The dominant energy condition reads 
$|w_A|\leq 1$ and $|w_B|\leq 1$.
This condition is violated 
in a neighborhood of the boundaries $\alpha = 0$ and
$\alpha = 1$ of the state space; to see this we observe that
\[
w_{A}\big|_{\alpha=0}=w+2>1\:,\qquad w_{A}\big|_{\alpha=1}=w-2<-1 \:.
\]
Hence, by continuity, there is a neighborhood of the sets 
$\alpha=0$ and $\alpha=1$ in the phase space where the dominant energy condition is violated,
see the shadowed region in 
Figure~\ref{lrspicture};
when $v \rightarrow 0$,  
this region becomes smaller and smaller
(and eventually reduces to the empty set for $v=0$).
Note that we only have 
directional dominant energy condition violation,
i.e., only one of the normalized principal pressures
(either $w_A$ or $w_B$) is bigger than one.
In particular, the isotropic pressure, $w = 1/3(w_A + 2 w_B)$
always satisfies $|w| <1$.
The violation of energy conditions is thus much milder than
in the context of phantom fields or similar models.
In contrast to energy condition violation toward the singularity, 
at late times the dominant energy condition is always satisfied.
In fact, for every solution, the energy conditions are satisfied
for all times larger than some given time. (For a sufficiently small elastic constant $v$, 
the energy conditions are satisfied already after Planck's time.)
Finally, let us draw the reader's attention to~\cite{CH2}, where 
a close connection between 
energy condition violation and oscillatory
singularities is established.

In general, it is unrealistic to expect that 
the description of the matter as an elastic material---represented 
by the constitutive equation of state~\eqref{rho}---remains true
under extreme stresses.
Under extreme conditions the material will loose its elastic properties
and its behavior might deviate considerably from the one described.
For instance, the assumption of a constitutive equation of state of
the quasi-Hookean form 
is typically justified only under the condition of small shear, see~\cite{KS},
and hence violated if the shear scalar $s^2$, cf.~\eqref{shearsca},
becomes too large; this occurs when the variable 
$\alpha$ is too close to $0$ or $1$ (which characterises
for the approach to the singularity). It is thus to be
expected that the description of a material as
elastic has a limited range of validity---it breaks down
before the singularity is reached.
Clearly, a modification of the material's properties under extreme
conditions will lead
to different asymptotic behavior of the associated cosmological
models. We refrain from investigating these issues 
further here since any modification of the matter properties 
seems rather ad hoc rather than based on sound physical considerations.
However, we refer to~\cite{CH2}, where we consider 
more general anisotropic matter sources.

The observed results on the 
long-term evolution of elastic cosmological models
are completely independent of the above considerations,
especially since
we study elastic matter whose material properties are 
almost indistinguishable from those of perfect fluids.
Let us reemphasize that we do not propose elastic matter as a 
physically realistic matter model for the universe;
however, the results 
bear relevance 
on our general understanding of the physics of cosmological models.
The
isotropization rates of elastic cosmological models
differ
from those of perfect fluid models even in the case where
the material properties of the elastic matter deviate
from those of perfect fluid matter by an arbitrarily small amount.
Whether the difference in isotropization rates
found here carries over to more general 
cosmological models with more general matter sources 
remains to be seen.

\vspace{0.5cm}

\noindent {\bf Acknowledgments:} 
We would like to thank an anonymous referee for valuable comments.
S.\ C.\ is supported by  Ministerio Ciencia e Innovaci\'on, Spain (Project MTM2008-05271).

\begin{appendix}

\section*{Appendix: Elasticity theory}\label{elasticapp}

This appendix is devoted to a presentation of some basic concepts of the general relativistic 
theory of elasticity. In particular, we specify in detail 
the assumptions on the elastic matter model used in this paper, which lead 
to the energy density~\eqref{rho} and the principal pressures~\eqref{wAB}.
A more detailed exposition of relativistic elasticity can be 
found in the references listed at the end of the paper. 

The {\it reference state} of an elastic body is defined to be the state of the body in the absence of strain and external forces. 
(The reference state has of course to be understood in a Platonic sense, since the conditions of vanishing strain 
and external forces cannot be realized in the the real world.) It is assumed that, in the
reference state, the particles of the body form, in the continuum limit, a smooth three 
dimensional manifold $N$, called {\it material space}. 
The 
manifold $N$ must be equipped with different tensor fields in order to describe the physical 
properties of the body in the reference state. The very least one has to require is the ability 
to ``count" the particles, and therefore to admit that the material space is equipped with a 
volume form (particle density). However, a consistent theory for the dynamics of elastic bodies 
in general relativity is presently available only under the stronger requirement that on the 
manifold $N$ be defined a Riemannian metric $\gamma$ (which reflects the ability to measure 
the ``distance" between the particles). We refer to~\cite{KS} for a discussion of the physical 
interpretation of this condition and on the restrictions that it imposes on the class of 
elastic materials covered by the theory. 

The state of the body in a four dimensional space-time $(M,\bar{g})$ is determined by a
\textit{configuration map} 
$\psi:M\to N$ with the property that for all $q\in N$ the set $\psi^{-1}(q)$ is a timelike curve 
(the world-line of the ``particle" $q$). This definition implies that the kernel of the \textit{deformation gradient} 
$T\psi: TM \rightarrow TN$ is generated by a (future-directed unit) timelike 
vector field $u$, which is interpreted as the matter four-velocity; by construction, $\psi^{-1}(q)$ is an integral curve.

In the following, let $x^\mu$ denote a system of local coordinates in the space-time and 
$X^A$ a system of local coordinates in the material space; Greek indexes run from 0 to 3, capital 
Latin indexes from 1 to 3 and refer to tensors on the material space. Let  $\psi^A(x^\mu)$ be the 
triple of functions representing the configuration map in these coordinates and $\partial_\mu\psi^A(x^\mu)$ 
the matrix-valued function representing the deformation gradient. Moreover we denote by 
$\gamma_{AB}$ the components of the material metric.
We define two metrics 
on the orthogonal complement $\langle u \rangle^\perp$ of $u$ in $TM$. The Riemannian metric induced by $\bar{g}$ we denote by $g$:
\[
g_{\mu\nu}=\bar{g}_{\mu\nu}+u_\mu u_\nu\,.
\]
The pull-back of the material metric by the map $\psi$, i.e., $\psi^*(\gamma)$, 
is called the \textit{relativistic strain tensor} $h$:
\begin{equation}\label{hdef}
h_{\mu\nu}=\partial_\mu\psi^A\partial_\nu\psi^B\,\gamma_{AB}\:.
\end{equation}
The relativistic strain tensor satisfies two fundamental properties:
\begin{itemize}
\item[(i)] It is orthogonal to the matter four-velocity, i.e., 
\begin{subequations}\label{hconst}
\begin{equation}
h_{\mu\nu}u^\mu=0\:;
\end{equation}
\item[(ii)] It is constant along the matter flow, i.e.,
\begin{equation}
\mathcal{L}_u h_{\mu\nu}=0\:.
\end{equation}
\end{subequations}
\end{itemize}
By the property (i), $h_{\mu\nu}$ defines a Riemannian metric on $\langle u \rangle^\perp$. Hence $h^\mu_{\ \nu}=g^{\mu\sigma}h_{\sigma\nu}$ 
has three positive
eigenvalues $h_1$, $h_2$, $h_3$. 

The material is unstrained at $x$ iff $g_{\mu\nu}(x)=h_{\mu\nu}(x)$. The scalar quantity
\[
n=\sqrt{{\rm det}_gh}=\sqrt{h_1h_2h_3}
\]
is the \textit{particle density} of the material. 
This interpretation is justified by virtue of the continuity equation 
\[
\nabla_\mu\left(nu^\mu\right)=0\:.
\]

A specific choice of elastic material is made 
by postulating a \textit{constitutive equation}, i.e., the 
functional dependence of the (rest frame) energy density $\rho$ of 
the material on the configuration map, the deformation gradient 
and the space-time metric. An important class of materials is the 
one for which this functional dependence enters only through 
the principal invariants of the strain 
tensor. In this case we have
\begin{equation}\label{lagrangian}
\rho=\rho(q_1,q_2,q_3)\:,
\end{equation}
where 
\[
q_1=\mathrm{tr}\, h\:,\qquad q_2=\mathrm{tr}\left(h^2\right),\qquad q_3=\mathrm{tr}\left(h^3\right);
\] 
since $n^2=(q_1^3-3q_1q_2+2q_3)/6$,
one of the invariants $q_i$ can be replaced by the particle density $n$.
The materials described by~\eqref{lagrangian} 
generalize the class of isotropic, homogeneous, 
hyperelastic materials from the classical theory of elasticity, 
see~\cite{MH}. As explained in~\cite{BS}, restriction 
to these materials guarantees 
that the resulting  elasticity theory is generally covariant and therefore 
seems quite a natural assumption in the context of general relativity. 

In this paper we make use of a constitutive equation introduced in~\cite{KS}. Let the \textit{shear scalar} to be defined as  
\begin{subequations}\label{shearsca}
\begin{align}
s^2 & =\frac{1}{36}\left[n^{-2}\left(q_1^3-q_3\right)-24\right]\,, 
\intertext{or, in terms of the eigenvalues $h_1$, $h_2$, $h_3$,} 
\label{shearscalar}
s^2 & =\frac{1}{12}\left[
\left(\sqrt{\frac{h_1}{h_2}}-\sqrt{\frac{h_2}{h_1}}\right)^2+
\left(\sqrt{\frac{h_1}{h_3}}-\sqrt{\frac{h_3}{h_1}}\right)^2+
\left(\sqrt{\frac{h_2}{h_3}}-\sqrt{\frac{h_3}{h_2}}\right)^2\right]\,.
\end{align}
\end{subequations}
Evidently, $s^2$ is non-negative, and $s^2=0$ (no shear) iff $h_{\mu\nu} \propto g_{\mu\nu}$ (or equivalently, $h_1 = h_2 = h_3$).
Following~\cite{KS} we shall consider a constitutive equation of the {\it quasi-Hookean} form, that is 
\begin{equation}\label{constitutiveeq}
\rho= \check{\rho}(n)+\check{\mu}(n) s^2\,,
\end{equation}
where $\check{\rho}(n)$ 
is the \textit{unsheared energy density} and $\check{\mu}(n)$ the \textit{modulus of rigidity} (or {\it shear modulus}).
The stress-energy tensor associated with these 
materials is obtained as the variation with respect to the 
space-time metric of the matter action $S_M=-\int\sqrt{|\bar{g}|}\,\rho$. 
The result is given in~\cite[Sec.~6]{KS} and reads
\begin{subequations}\label{Tmunu}
\begin{align}
\label{stressenergytensor}
\bar{T}_{\mu\nu}=\rho\, u_{\mu} u_{\nu} \, +\:\, & T_{\mu\nu}\,, \\[1ex]
\label{anisotropicstress}
\text{where}\quad\:
T_{\mu\nu}\, = \:\, & p\,\,g_{\mu\nu}+
\frac{1}{6}\frac{\check{\mu}}{n^2}
\left[\frac{1}{3}\left({\rm tr}(h^3)-({\rm tr} h)^3\right)g_{\mu\nu}+({\rm tr}h)^2h_{\mu\nu}-(h^3)_{\mu\nu}\right].
\end{align}
\end{subequations}
Here $p$ is the isotropic (component of the) pressure, which is given by
\begin{equation}\label{isopressure}
p=\check{p}(n)+ \check{\nu}(n) s^2\,, \qquad \text{where}\quad 
\check{p}=n^2\frac{d}{dn}\left(\frac{\check{\rho}}{n}\right)\,,\quad
\check{\nu} = \left(n\frac{d\check{\mu}}{dn}-\check{\mu}\right)\,.
\end{equation}
The principal pressures $p_i$ (which are the [non-zero] eigenvalues of $T^\mu_{\ \nu}$)
are thus of the form $p_i = p + \delta p_i$; for an unstrained configuration, $p_i = p$, $i=1,2,3$.
For $\check{\mu}=0$ (or $s^2=0$), 
the elastic material reduces to a perfect fluid 
with stress-energy tensor $\bar{T}_{\mu\nu} = \rho u_\mu u_\nu + p g_{\mu\nu}$,
energy density $\rho = \check{\rho}$ and pressure $p =\check{p}$.

It remains to specify the functions $\check{\rho}$ and $\check{\mu}$ 
in the constitutive equation~\eqref{constitutiveeq}. 
Following~\cite{KS}, we postulate a linear equation 
of state between the modulus of rigidity $\check{\mu}$ and the unsheared pressure $\check{p}$,
\begin{equation}\label{eqs1}
\check{\mu}  =v\,\check{p}\:, 
\end{equation}
where $v$ is a dimensionless constant that we call {\it elastic constant} and 
that is allowed to vanish (in which case the elastic body becomes a perfect fluid). Finally
we postulate 
a linear equation of state between the unsheared pressure $\check{p}$ and the unsheared energy density $\check{\rho}$,
\begin{equation}\label{eqs2}
\check{p}  = w\check{\rho}\:,
\end{equation}
where the constant $w$ is allowed to take values in the interval $[-1,1]$. We also assume that the product $vw$ is 
non-negative to guarantee the non-negativity of the energy density $\rho$ (see Eq.~\eqref{rhop} below). 
We remark that our choice of equation of state~\eqref{eqs2} is different from the one in~\cite{KS}; 
in ref.~\cite{KS}, the unsheared pressure and the unsheared energy density are assumed to satisfy a 
polytropic equation of state. The equation of state $\check{p}=\check{p}(\check{\rho})$ selects the 
perfect fluid model that arises as a special case of the elastic matter model when the modulus of 
rigidity vanishes. In our case, it is a perfect fluid that obeys a linear equation of state, 
which is the perfect fluid model most widely used in cosmology. In~\cite{KS}, 
where the applications concern the stellar dynamics case, the elastic body becomes
a polytropic perfect fluid star in the absence of rigidity.

Substituting~\eqref{eqs1} and~\eqref{eqs2} in~\eqref{isopressure} we find
\[
\check{\rho}  = \rho_0 n^{w+1}\,, 
\qquad \check{\mu}= \rho_0 vw\, n^{w+1}  \qquad (|w| \leq 1, \: vw \geq 0)
\]
for some constant $\rho_0>0$.
Accordingly,
\begin{equation}\label{rhop}
\rho=\rho_0 n^{w+1}\left(1+vw\, s^2\right),\qquad\qquad p=w\rho\,.
\end{equation}
Since for an unstrained material $\rho = \check{\rho}$ and
$p_i = p = \check{p}$ hold, $i=1,2,3$, the bound $|w| \leq 1$ ensures that the 
dominant energy condition $|p_i| \leq \rho$
is satisfied for an unstrained configuration.
Furthermore, $vw \geq 0$ guarantees that the energy density is positive for 
all values of the shear scalar $s^2$ and 
has a minimum at zero shear.
When $v = 0$, the modulus of rigidity $\check{\mu}$ vanishes and the elastic matter
reduces to a perfect fluid with a linear equation of state $p = w \rho$;
the condition $|w| \leq 1$ ensures that the dominant energy condition $|p| \leq \rho$
is satisfied for this perfect fluid.
When $w=0$ (so that $p = 0$), the choice of $v$ is irrelevant, since $vw = 0$; 
this is clear because shear cannot occur for dust.
Henceforth, unless stated otherwise, 
by elastic matter we will always mean matter with constitutive equation~\eqref{rho},
where $w \in [-1,1]$ and $vw \geq 0$. 

Consider now a homogeneous space-time $(M,\bar{g})$ of Bianchi type I, i.e.,
\begin{equation}\label{BianchiI}
\bar{g}_{\mu\nu} dx^\mu dx^\nu =-dt^2+g_{ij}(t)dx^idx^j\:,
\end{equation}
where $g_{ij}(t)$, $i,j=1,2,3$, is a family of Riemannian metrics
that is induced on the spatially homogeneous hypersurfaces $t=\mathrm{const}$. The spatial coordinates are 
chosen so that the Killing vector fields of the space-time coincide with the operators $\partial_{x^i}$. 
Now we recall that a matter model in a spatially homogeneous space-time is said to be {\it non-tilted} 
if the matter four-velocity $u$ is orthogonal to the hypersurfaces of spatial homogeneity. 
For the metric~\eqref{BianchiI} this means that $u\equiv\partial_t$. In this paper we assume that the 
elastic material in the Bianchi type~I space-time is non-tilted. By~\eqref{hconst} this implies
\begin{equation}\label{htimeind}
\partial_th_{\mu\nu}=0\:,\qquad h_{0\mu}=0\:,
\end{equation}
in particular the relativistic strain tensor is time independent. Next we recall (see, e.g.,~\cite{WE})
that compatibility with the Einstein equations entails that a matter model with stress-energy tensor $T_{\mu\nu}$ 
in a Bianchi type~I space-time must satisfy $\partial_{x^i}T_{\mu\nu}=0$, i.e. the stress- energy tensor is 
independent of the spatial variables. A matter model is said to inherit the Bianchi type~I symmetry if the 
relation $\partial_{x^i}T_{\mu\nu}=0$ implies that the matter fields, which enter into the definition of $T_{\mu\nu}$, 
are also independent of the spatial variables. For instance, in the case of perfect fluids, the 
relation $\partial_{x^i}T_{\mu\nu}=0$ implies the relations $\partial_{x^i}\rho=\partial_{x^i}p=0$ 
on the energy density and the pressure. In this paper we restrict to elastic matter that inherits 
the Bianchi type~I symmetry, in particular we assume that 
\begin{equation}\label{hindx}
\partial_{x^i}h_{\mu\nu}=0\:.
\end{equation} 
The equations~\eqref{htimeind} and~\eqref{hindx} imply that the component of the relativistic strain tensor are constant:
\begin{equation}\label{hfin}
h_{00}=h_{0k}=0,\qquad h_{ij}= \mathrm{const}\:.
\end{equation} 
These constants can be fixed arbitrarily and correspond to the ``initial data" for the matter field. 
Note that the relativistic strain tensor has been fixed by our geometric assumptions and that the concept 
of material space and material metric become redundant. (This is of course a consequence of the high 
degree of symmetry of the space-time). However at this point it is instructive to derive a material 
metric and configuration maps that give rise to a relativistic strain tensor of the form~\eqref{hfin}. 
We may consider a flat material metric and fix coordinates on the material space such that $\gamma_{AB}=\delta_{AB}$; 
then we restrict to homogeneous deformations (see~\cite{MH})
\[
\partial_t\psi^A=0\:,\quad\partial_{x^i}\psi^A=F_i^{\ A}=\mathrm{const}\quad\Rightarrow\quad \psi^A=F_i^{\ A}x^i+c^A\:,
\]
where $c^A$ is a constant.
By~\eqref{hdef} we obtain the relativistic strain tensor~\eqref{hfin}, where $h_{ij}=\delta_{AB}F_i^{\ A}F_j^{\ B}$.
Substituting~\eqref{hfin} in~\eqref{Tmunu} we obtain
\begin{equation}\label{T00ij}
\bar{T}_{00}=\rho,\qquad\bar{T}_{0k}=j_k= 0,\qquad \bar{T}_{ij}=T_{ij}\,
\end{equation}
where $T_{ij}$ is given in terms of $h_{ij}$ via~\eqref{anisotropicstress}.

The results presented in this paper have been obtained for {\it diagonal} solutions of the Einstein-elastic 
matter equations, i.e., solutions for which the space-time metric and the stress-energy tensor are diagonal. 
These solutions correspond to initial data, say at time $t=0$, of the following type. Let $g_{ij}(0)$ be the 
initial spatial metric and $k_{ij}(0)$ the second fundamental form of the hypersurface $t=0$. Without loss of 
generality we can assume that $g_{ij}(0)$ and $k^i_{\ j}(0)$
are diagonal 
(by choosing coordinates adapted to an orthogonal basis of eigenvectors of $k^i_{\ j}(0)$).
Furthermore we impose the condition that $h_{ij}$ is diagonal; 
in particular, by rescaling the spatial coordinates, we can assume $h_{ij}=\delta_{ij}$. 
The Einstein equations, in units $c = 1 = 8 \pi G$,
decompose into the momentum constraint $j_k =0$, 
which is automatically satisfied by~\eqref{T00ij}, 
the Hamiltonian constraint
\begin{subequations}\label{einstein}
\begin{equation}\label{constraint}
(\mathrm{tr}k)^2-k^i_{\ j}k^j_{\ i}-2\rho=0\:,
\end{equation}
and the evolution equations
\begin{equation}\label{evolution}
\partial_t g_{ij}=-2k_{ij}\quad\partial_{t}k^i_{\ j}=({\rm tr}k)k^i_{\ j}-T^i_{\ j}+\frac{1}{2}\delta^i_{\ j}(T^k_{\ k}-\rho)\:.
\end{equation}
\end{subequations}
Since the off-diagonal elements of the tensor $T^i_{\ j}$ form an homogeneous polynomial 
in $h^i_{\ j} = g^{i k} h_{j k}$, $i\neq j$, 
it follows from the evolution equations~(\ref{evolution}) 
that $(g_{ij}, k^i_{\ j}, h^i_{\ j})$ (and therefore $T^i_{\ j}$) remain diagonal for all times. 
We refer
to this special class of solutions of the equations~\eqref{einstein}
as \textit{diagonal models}. 

From $h^i_{\ j}= g^{ik} h_{k j} = \mathrm{diag}(g^{11},g^{22},g^{33})=\mathrm{diag}(h_1,h_2,h_3)$ 
we conclude that
\begin{equation}
s^2=\frac{1}{12}
\left[\frac{g^{11}}{g^{22}}+\frac{g^{22}}{g^{11}}+\frac{g^{11}}{g^{33}}+\frac{g^{33}}{g^{11}}+\frac{g^{22}}{g^{33}}+\frac{g^{33}}{g^{22}}-6\right],
\end{equation}
cf.~(\ref{shearscalar}), which can be inserted into~\eqref{rhop}, i.e.,
\begin{subequations}\label{stressE}
\begin{equation}\label{rhoagain}
\rho = \rho_0\, (g^{11} g^{22} g^{33})^{(w+1)/2} \,( 1 + vw s^2)\,,
\end{equation}
to yield $\rho$ as a function of $g^{11}$, $g^{22}$, $g^{33}$.
Moreover, from~(\ref{anisotropicstress}) we find
\begin{align}
T^1_{\ 1}& =p+\frac{1}{6}\check{\mu}\left(\frac{g^{11}}{g^{33}}-\frac{g^{33}}{g^{11}}+\frac{g^{11}}{g^{22}}-\frac{g^{22}}{g^{11}}\right),\\
T^2_{\ 2}& =p+\frac{1}{6}\check{\mu}\left(\frac{g^{22}}{g^{11}}-\frac{g^{11}}{g^{22}}+\frac{g^{22}}{g^{33}}-\frac{g^{33}}{g^{22}}\right),\\
T^3_{\ 3}& =p+\frac{1}{6}\check{\mu}\left(\frac{g^{33}}{g^{22}}-\frac{g^{22}}{g^{33}}+\frac{g^{33}}{g^{11}}-\frac{g^{11}}{g^{33}}\right),
\end{align}
\end{subequations}
where $p = w \rho$ and $\check{\mu} = \rho_0 vw (g^{11}g^{22}g^{33})^{(w+1)/2}$ and are thus functions of $g^{11}$, $g^{22}$, $g^{33}$.

Finally, in the local rotational symmetry case, where we assume that the plane of 
rotational symmetry is the $x_2$-$x_3$ plane, the metric takes the form~\eqref{metric}. 
Substituting $g^{11}=1/A$ and $g^{22}=g^{33}=1/B$ in Eq.~\eqref{rhoagain} we obtain the 
expression~\eqref{rho}  for the energy density. Furthermore, defining $p_A=T^1_{\ 1}$, $p_B=T^2_{\ 2}$, 
the rescaled principal pressures $w_A=p_A/\rho$, $w_B=p_B/\rho$  are given by~\eqref{wAB}.

\end{appendix}



%




%












\end{document}